\documentclass[aps,preprint,nofootinbib,superscriptaddress,prc]{revtex4-1}

\usepackage{epsfig}
\usepackage{bm}
\usepackage[bookmarksnumbered,bookmarksopen,colorlinks,citecolor=blue,linkcolor=blue] {hyperref}
\begin{document}


\title{Isospin equilibration in multinucleon transfer reaction at near-barrier energies}


\author{Cheng Li}\email{licheng@mail.bnu.edu.cn}

\author{Cheikh A.T. Sokhna}
\author{Xinxin Xu}
\author{Jingjing Li}
\author{Gen Zhang}
\author{Bing Li}
\author{Zhishuai Ge}
\affiliation{Beijing Radiation Center, Beijing 100875, China}
\affiliation{The Key Laboratory of Beam Technology and Material Modification of Ministry of Education, College of Nuclear Science and Technology, Beijing Normal University, Beijing 100875, China}

\author{Feng-Shou Zhang$^{1,2,}$}
\email[Corresponding author, ]{fszhang@bnu.edu.cn}
\affiliation{Center of Theoretical Nuclear Physics, National Laboratory of Heavy Ion Accelerator of Lanzhou, Lanzhou 730000, China}

\begin{abstract}
The isospin equilibration process in multinucleon transfer reaction is investigated by using the improved quantum molecular dynamics model. The collision processes of $^{124}$Xe+$^{208}$Pb at near-barrier energy are studied with different symmetry energy coefficients. We find that neutrons transfer happens earlier than protons. The large symmetry energy promote the transfer of neutrons. The neutron flow from the target to projectile is along the low-density path of neck. The isospin equilibration process in $^{58}$Ni+$^{208}$Pb reaction is also investigated and compared with available experimental data. It shows that $N/Z$ values of the projectile-like products increase rapidly with increasing mass transfer. The complete isospin equilibration events are located in the region of $120<A<150$ which are produced in symmetric fission-type reactions.
\end{abstract}

\maketitle
\begin{center}
\textbf{I. INTRODUCTION}
\end{center}
The isospin transport phenomenon occurs in collisions between the projectile and the target with different $N/Z$ ratio \cite{iso1,iso2,iso3,iso4,iso5,iso6,iso7,iso8}. It will drive the system toward a uniform asymmetry distribution. When two nuclei with different $N/Z$ asymmetries come into contact, the isospin transport is initiated and continues until the system disintegrates or the chemical potentials for neutrons and protons in both nuclei become equal. If the interaction time between the projectile and target is long enough, the system will reach the status of isospin equilibration. The interaction time depends on the reaction partners, incident energies, and impact parameter of the collision. For intermediate energies or in peripheral heavy-ion collisions, the reaction partners may interact for a short time, and exchange of particles between the reaction partners does not lead to a uniform $N/Z$ distribution through the whole system. The multinucleon transfer (MNT) reactions at near-barrier energies have attracted widespread interests in recent years both experimentally \cite{exp1,exp2,exp3,exp4,exp5,exp6,exp7,exp8,exp9,exp10} and theoretically \cite{rew1,rew2,rew3,rew4,rew5,rew6,rew06,rew7,rew8,rew9,rew10,rew11,rew12}. For a comprehensive review of MNT reactions, one can see the Ref. \cite{exp11}. The interaction time in the MNT reactions could be extraordinarily long. Especially in the quasifission collisions, the typical time scales can reach $10^{-21}$ s order of magnitude or longer \cite{quas1,quas2}. In this situation, the nucleon exchange processes may lead to a uniform distribution of the $N/Z$ asymmetry.

The isospin equilibration effect in the reactions of $E_{\textmd{lab}}=275$ MeV $^{64}$Ni with $^{130}$Te and $E_{\textmd{lab}}=345$ MeV $^{58}$Ni with $^{208}$Pb has been investigated by Kr\'{o}las et al. at Laboratori Nazionali di Legnaro (INFN) \cite{exp7}. It provides a strong evidence that the isospin equilibration process is closely related to the interaction time of the reaction partners. The isospin equilibration process in the MNT reactions is also an important path to produce new neutron-rich nuclei. For example, the new isotopes $^{54}$Ti, $^{56}$V, $^{58,59}$Cr, $^{61}$Mn and $^{63,64}$Fe were produced in 1980 by Guerreau et al. through a 340 MeV $^{40}$Ar beam accelerated by the Orsay ALICE accelerator facility bombarding on $^{238}$U target \cite{new1}. In addition, the neck formation is an important characteristic during the evolution process of the system at low-energy heavy-ion collisions \cite{rew9}. The isospin transport in the neck can produce some interesting phenomena. Such as the neutron transfer between the target and the projectile will reduce the effective barrier and enhance the fusion cross sections at sub-barrier energies \cite{bar1}.

The dynamical calculation is required to understand some details of isospin equilibration process. The microscopic dynamics models such as the time-dependent Hartree-Fock (TDHF) model \cite{TDHF1,TDHF2,TDHF3,TDHF4,TDHF5,TDHF6} and the improved quantum molecular dynamics (ImQMD) model \cite{QMD1,QMD2,QMD3,QMD4,QMD6} describe the nuclear reactions based on mean-field approximation derived from an effective nucleon-nucleon interaction. Both the TDHF and ImQMD models are self-consistent models for describing the behavior of neck growth and nucleon transport during the collisions. S. Ayik et al. investigated the nucleon exchange mechanism in $^{40}$Ca+$^{238}$U and $^{48}$Ca+$^{238}$U systems by using the TDHF model \cite{TDHF2}. The calculations show that the drift coefficient of the neutron is larger than that of proton during the collisions in both $^{40}$Ca+$^{238}$U and $^{48}$Ca+$^{238}$U systems. In the ImQMD model, the standard Skyrme interaction with the omission of the spin-orbit term is adopted for describing the bulk properties and the surface properties of nuclei \cite{QMD4}. The stochastic two-body collision process is added to the time evolution by the Hamilton equation of motion. The final state of the two-body collision process is checked so that it obeys the Pauli principle. The ImQMD model is successfully applied to heavy-ion fusion reactions at energies near the Coulomb barrier \cite{QMD2,QMD3} and intermediate energy heavy-ion collisions \cite{QMD4,QMD6}. In this work, we apply the ImQMD model to investigate the isospin equilibration process in the reactions of $^{124}$Xe+$^{208}$Pb and $^{58}$Ni+$^{208}$Pb at near-barrier energies.

The structure of this paper is as follows. In Sec. II, we briefly introduce the ImQMD model. The results and discussion are presented in Sec. III. Finally, the conclusion is given in Sec. IV.

\begin{center}
\textbf{II. THE MODEL }
\end{center}
The ImQMD model is an improved version of the quantum molecular dynamics (QMD) model \cite{QMD1} which takes into account the effects of the surface term, the phase-space density constraint, the surface-symmetry potential term, and so on. The nuclear interaction potential energy $U_{\textrm{loc}}$ is obtained from the integration of the Skyrme energy density functional $U=\int V_{\textrm{loc}}(\mathbf{r})d\mathbf{r}$ without the spin-orbit term, which reads
\begin{eqnarray}
\nonumber V_{\mathbf{loc}}=&&\frac{\alpha }{2}\frac{\rho ^{2}}{\rho _{0}}+\frac{\beta }{\gamma +1}%
\frac{\rho ^{\gamma +1}}{\rho _{0}^{\gamma }}+\frac{\textsl{g}_{sur}}{2\rho _{0}}%
(\nabla \rho )^{2}\\
&&\ +\frac{C_{s}}{2\rho _{0}}(\rho ^{2}-\kappa _{s}(\nabla \rho
)^{2})\delta ^{2} + g_{\tau}\frac{\rho ^{\eta +1}}{\rho_{0}^{\eta
}}. \label{12}
\end{eqnarray}
Here $\rho=\rho_{n}+\rho_{p}$ is the nucleons density. $\delta=(\rho_{n}-\rho_{p})/(\rho_{n}+\rho_{p})$ is the isospin asymmetry. The density distribution function $\rho$ of a system can be read \begin{equation}
\rho(r)=\sum_{i}\frac{1}{(2\pi\sigma_r)^{3/2}}\exp[-\frac{(\bm r-\bm r_i)^2}{2{\sigma_r}^2}]. \label{aba:app1}
\end{equation}
$\sigma_r$ is the wave-packet width of the nucleon in coordinate space. The IQ2 parameter sets (see Table 1) adopted in this work are the same as in Refs. \cite{rew2,rew9,QMD6}. The incompressibility coefficient, $K_\infty$, is 195 MeV. This parameter sets have been successfully applied on the heavy-ion collisions in fusion reactions, multinucleon transfer reactions, and ternary breakup reactions. We have adopted the Fermi constraint to describe the fermionic nature of the $N$-body system which improve greatly the stability of an individual nucleus. The two-body collision correlations and the Pauli blocking checking are also included \cite{Papa}.

\begin{table}[h]
\tabcolsep=1pt \caption{The model parameters (IQ2) adopted in this work.}
{\begin{tabular}{@{}cccccccccc@{}}

\hline\hline
   $\alpha$ & $\beta$ & $\gamma$ & $\textsl{g}_{sur}$ & $\textsl{g}_{\tau}$ & $\eta$ & $C_{S}$ & $\kappa_{s}$ & $\rho_{0}$ \\
  $(\textrm{MeV})$ & $(\textrm{MeV})$ & $$ & $(\textrm{MeV}~ \textrm{fm}^{2})$ & $(\textrm{MeV})$ & $$ & $(\textrm{MeV})$ & $(\textrm{fm}^{2})$ & $(\textrm{fm}^{-3})$ \\
\hline
   $-356$ & $303$ & $7/6$ & $7.0$ & $12.5$ & $2/3$ & $32.0$ & $0.08$ & $0.165$\\
\hline\hline
\end{tabular}}

\end{table}

In this work, we set $z$-axis as the beam direction and $x$-axis as the impact parameter direction. We use the parameter sets of IQ2 and set the wave-packet width $\sigma_r=1.3$ fm to calculate the isospin equilibration process of $^{124}$Xe+$^{208}$Pb and $^{58}$Ni+$^{208}$Pb. The initial distance of the center of mass between the projectile and target is 30 fm.

We first test the ImQMD model for the description of MNT reactions. The collisions of $^{58}$Ni+$^{208}$Pb at $E_{\textrm{lab}}=328.4$ MeV are simulated by using the ImQMD model. 39000 simulation events are calculated for the impact parameters, $b = 1$, 2, 3...13 fm. For each event, we simulate the whole collision process until $t=2000$ fm/c with a step size of $t= 1$ fm/c. Fig. 1 shows the isotope production cross sections from Mn to Ni. The thick folding lines and thick solid lines denote the calculation results from the combination of ImQMD+GEMINI and GRAZING model with inclusion of the evaporation. From Fig. 1, one can see that the GRAZING model \cite{Grazing1,Grazing2,Grazing3} estimates the production cross sections very well at 0p and -1p channels. It grossly underestimates the production cross section in the case of more proton transfers. The production cross sections from Mn to Ni calculated by using the combination of ImQMD+GEMINI are in good agreement with the experimental data \cite{Ni1}. The nuclear level densities in the GEMINI code \cite{gemini} are taken as a Fermi-gas form with the default parameters. A good descriptions of the ImQMD+GEMINI model in the isotopic production cross sections can also be found in the calculations of $^{136}$Xe+$^{208}$Pb \cite{rew9} and $^{136}$Xe+$^{198}$Pt \cite{rew2} systems.

\begin{figure*}[!htb]
\includegraphics[width=0.8\hsize]{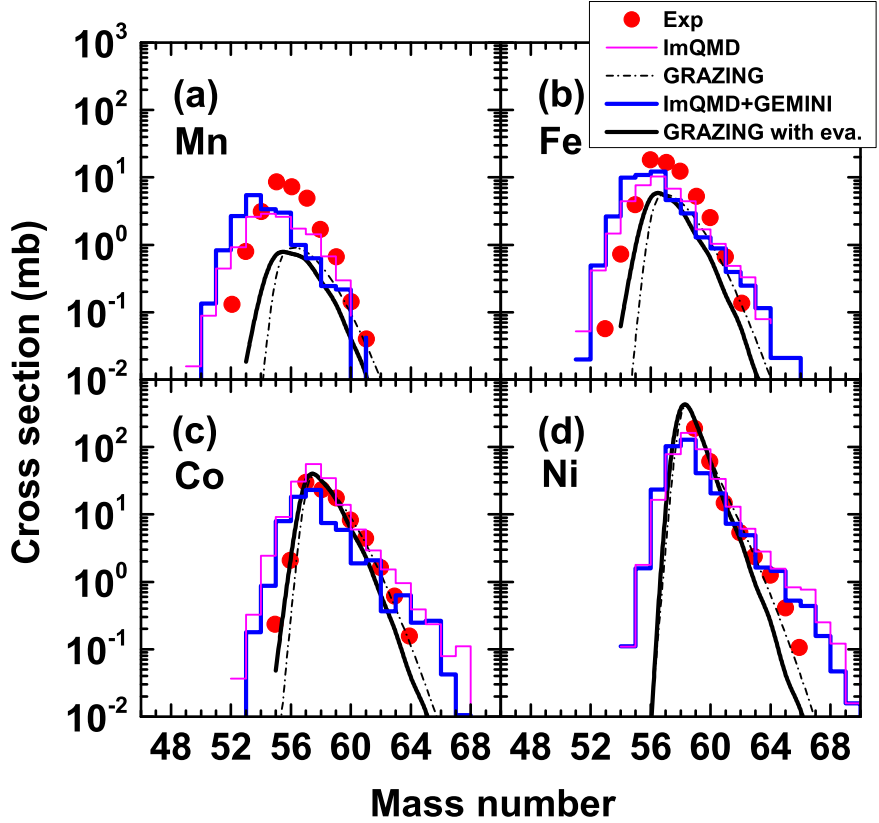}
\caption{(Color online.) Isotope production cross sections from Mn to Ni in the reaction of $^{58}$Ni+$^{208}$Pb at $E_{\textmd{lab}}=328.4$ MeV. The thick folding lines and thick solid lines denote the calculation results from the combination of ImQMD+GEMINI and GRAZING model with inclusion of the evaporation. The thin folding lines and thin dash-dotted lines denote the primary fragment distributions from the ImQMD and GRAZING model, respectively. The experimental data (solid circles) are from the Ref. \cite{Ni1}.}
\label{fig1}
\end{figure*}

\begin{center}
\textbf{III. RESULTS AND DISCUSSION}
\end{center}
For simplicity, we consider head-on collisions of $^{124}$Xe+$^{208}$Pb at $E_{\textrm{c.m.}}=450$ MeV to investigate the isospin equilibration process. Firstly, we show the single-particle potential of $^{124}$Xe+$^{208}$Pb along the beam direction at $t=200$ and 300 fm/c in Fig. 2. The projectile and target in the initial state are located on the left and right side, respectively. The single-particle potential is defined as $V_{sp}(\textbf{r})=\int \rho(\textbf{r}^\prime)V(\textbf{r}-\textbf{r}^\prime)d\textbf{r}^\prime$, where $\rho(\textbf{r})$ is the density distribution of the system and $V(\textbf{r}-\textbf{r}^\prime)$ is the effective nucleon-nucleon interaction. The ImQMD calculations show that the value of the neutron potential in inner of two nuclei is about -35 MeV. Due to the Coulomb interaction, the proton potentials in inner of two nuclei are higher (about -15 MeV) than the neutron potentials. In addition, there exists a higher barrier at the edge of the nuclei. This is a very important characteristic in the nuclear reactions at near-barrier energies. For example, at $t=200$ fm/$c$, the neutron transfer path is opened because the inner potential barrier between the two reaction partners is negative (see Fig. 2(a)). However, for protons, there is a high enough inner potential barrier which prevents the protons transfer from the target to projectile (see Fig. 2(c)). Hence, when the projectile and target get close to each other, the neutrons transfer happens earlier than the protons. With growing up of the neck, the inner potential barriers for neutrons and protons are reduced. We can see that the inner barrier for neutrons and protons are about -25 and -5 MeV at $t=300$ fm/$c$, respectively. In this situation, both the neutrons and protons are allowed to transfer between the target and projectile (see Fig. 2(b) and 2(d)).
\begin{figure*}[!htb]
\includegraphics[width=0.7\hsize]{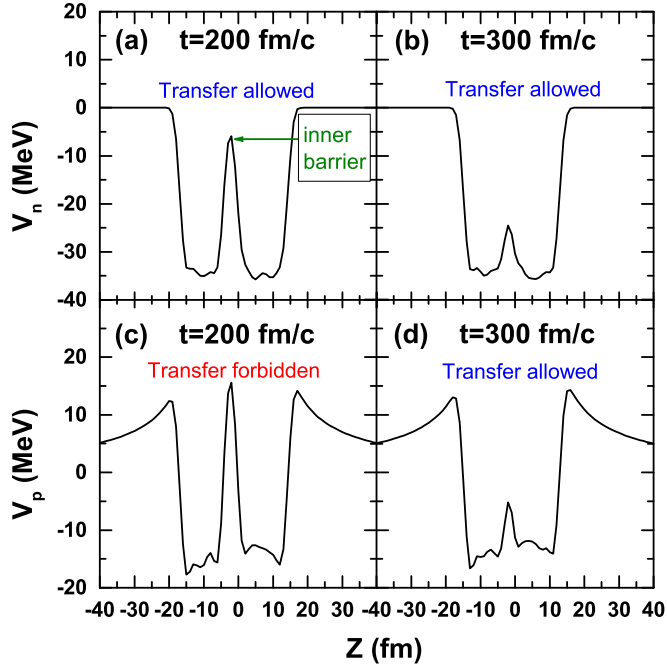}
\caption{(Color online.) The neutron (up panels) and proton (bottom panels) potential for $^{124}$Xe+$^{208}$Pb reaction at $t=200$ and $t=300$ fm/$c$, respectively.}
\label{fig2}
\end{figure*}

Fig. 3 shows that the isospin asymmetry ($\delta=(\rho_{n}-\rho_{p})/(\rho_{n}+\rho_{p})$) distributions for the $^{124}$Xe+$^{208}$Pb system during the evolution of the reaction. The average isospin asymmetry for $^{124}$Xe and $^{208}$Pb are 0.13 and 0.21, respectively. From Fig. 3, one sees that the structure effect on the isospin asymmetry is very notable. The isospin asymmetry in the inner of two reaction partners is uniform. The values of $\delta$ for the projectile-like and target-like nuclei are 0.1 and 0.17, respectively. However, the isospin asymmetry on the surface is significantly larger compared to the inner of the nuclei. It can be seen that the core of $^{208}$Pb is covered by neutron skin with maximal $\delta=0.54$ at $t=50$ fm/c. This is because that for a neutron-rich system, the density corresponding to the minimum of the chemical potential of neutrons is lower than that of protons \cite{chem}. Hence, the neutrons are preferably driven to the low density area. When the projectile and target contact to each other ($t=200$ fm/c), the neutrons gathered in the low density overlap area. The neutron flow between the projectile and target can be found, which is along the low density path of the neck.
\begin{figure*}[!htb]
\includegraphics[width=0.9\hsize]{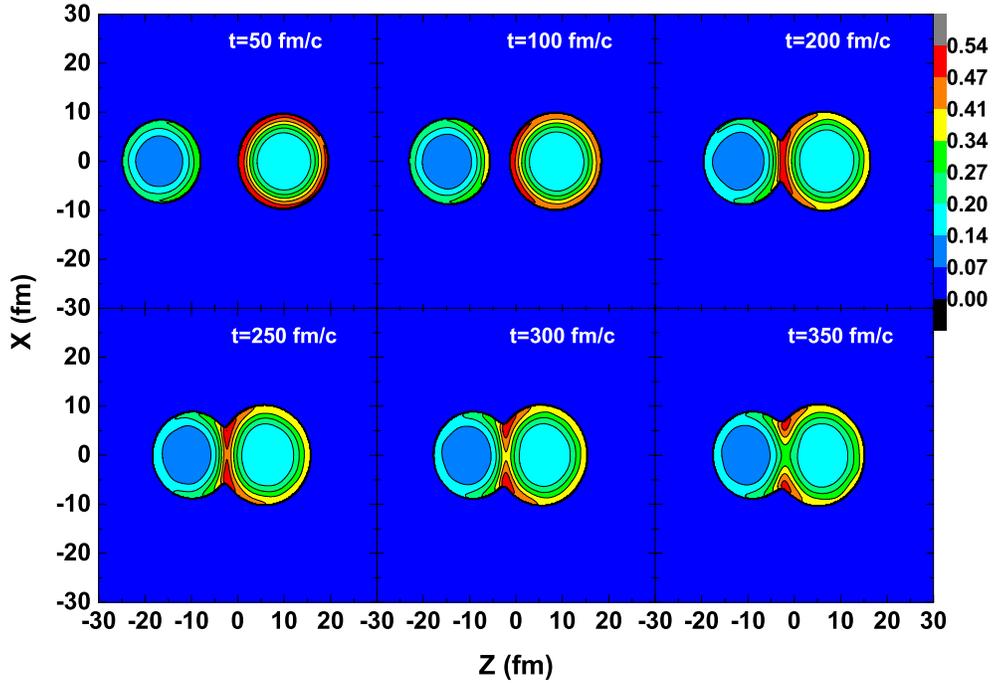}
\caption{(Color online.) The isospin asymmetry $\delta=(\rho_{n}-\rho_{p})/(\rho_{n}+\rho_{p})$ distribution of $^{124}$Xe+$^{208}$Pb with the time evolution.}
\label{fig3}
\end{figure*}

\begin{figure*}[!htb]
\includegraphics[width=0.7\hsize]{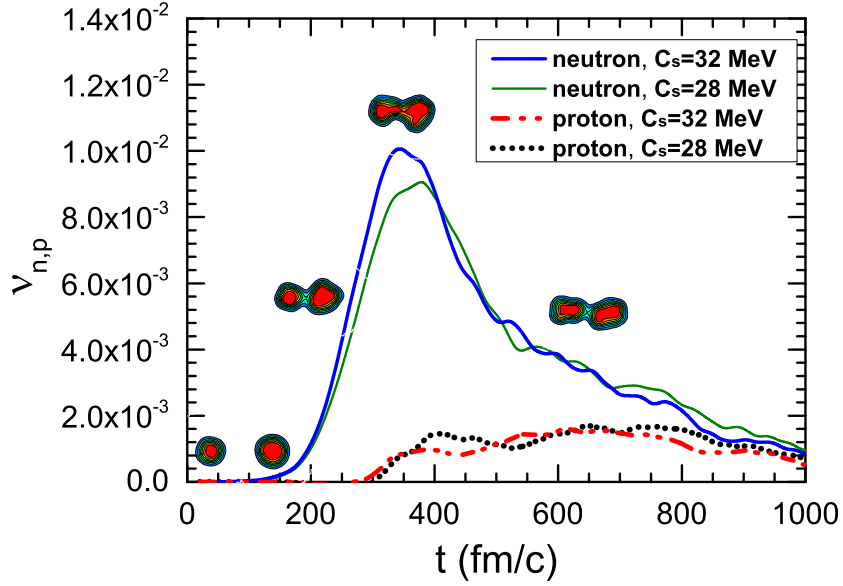}
\caption{(Color online.) The drift coefficients for neutron and proton with symmetry energy coefficients $C_s=28$ and 32 MeV in the central collisions of $^{124}$Xe+$^{208}$Pb system at incident energy $E_{\textrm{c.m.}}=450$ MeV. The contour plots in the figure are the density distributions of the system at different reaction stage.}
\label{fig4}
\end{figure*}

In the head-on collisions of $^{124}$Xe+$^{208}$Pb, the incident energy in the center-of-mass frame is slightly higher than the Coulomb barrier (436 MeV). The neck evolution is an important characteristic, which includes contact, growing up, and re-separation processes. In order to investigate the nucleon transfer with the evolution of the neck, we define the separation plane of projectile-like and target-like nuclei at the position where iso-contours of the projectile and target densities cross each other. This method was also adopted by the TDHF calculation in Refs. \cite{den1,den2}. Fig. 4 shows the drift coefficients of neutron and proton in the central collisions of $^{124}$Xe+$^{208}$Pb system with the symmetry energy coefficients $C_s=28$ and 32 MeV. The drift coefficient can be read as $\upsilon_{n,p}=dN_{n,p}/dt$, where $N_{n,p}$ denotes the net neutron or proton flux through the separation plane. The drift direction of nucleons is from the target to projectile. It can be seen that the neutron transfer starts at $t=150$ fm/c due to a lower neutron potential barrier between the reaction partners at the touching configuration. The neutron transfer rapidly increases with the growing up of the neck. Due to the stronger symmetry potential, the neutron transfer in the case of $C_s=32$ MeV is faster compared to the case of $C_s=28$ MeV. The peak value of the drift coefficient in the case of $C_s=32$ MeV is about $0.3\times10^{22}\textrm{s}^{-1}$ which is very close to the TDHF calculation in Ref. \cite{TDHF2}. At the re-separation process of the system, the neutron drift coefficients decrease with the disappearing of the neck. The protons transfer starts at $t=300$ fm/c. However, we note that the proton drift coefficients are weak correlation to the symmetry potential intensity.

\begin{figure*}[!htb]
\includegraphics[width=0.8\hsize]{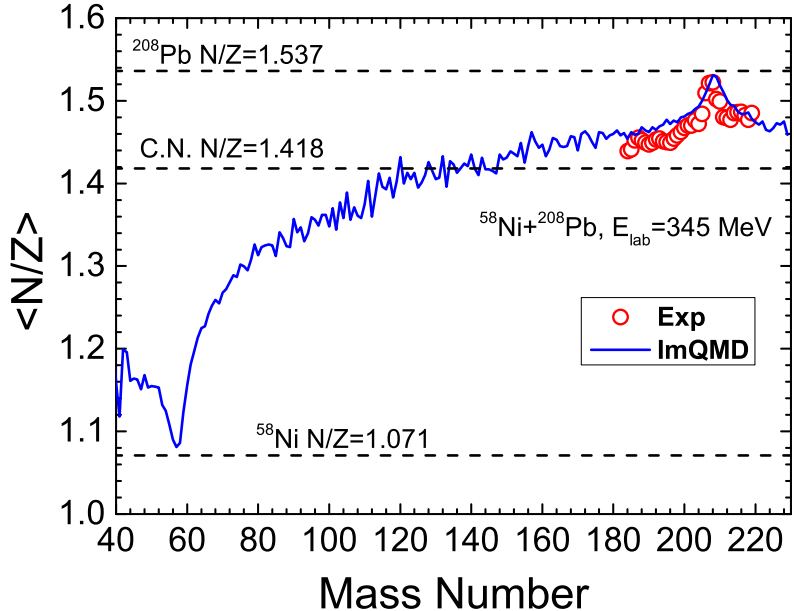}
\caption{(Color online.) Average $N/Z$ values for primary products of $^{58}$Ni+$^{208}$Pb at $E_{\textrm{lab}}=345$ MeV. The solid line is the calculation of the ImQMD model. The dashed lines indicate the $N/Z$ ratios of the target, the projectile and the compound system. The fusion-fission events are not included in primary products. The experimental data (open circles) are reconstructed primary fragments taken from the Ref. \cite{exp7}.}
\label{fig5}
\end{figure*}

In order to compare with experimental results, we show the average $N/Z$ values for primary products of $^{58}$Ni+$^{208}$Pb at $E_{\textrm{lab}}=345$ MeV in Fig. 5. One sees that the experimental data can be reproduced very well by the ImQMD calculation. The $N/Z$ values of products near the projectile and target are close to the initial values, i.e., 1.071 for the projectile and 1.537 for the target. With smaller mass transfer (about 5 mass units), the $N/Z$ values of the projectile-like products increase rapidly due to the transferring of more neutrons from the target in the quasi-elastic collisions. Hence, one can see that a steep valley appears in the position of the projectile, and corresponding a peak appears in the position of the target. The interaction time in the quasi-elastic reactions is very short. The system can't reach the full isospin equilibration. The complete isospin equilibration fragments (located in the region of $120<A<150$) are produced in symmetric fission-type reactions. This interesting feature can also be found in the reactions of $^{64}$Ni+$^{130}$Te \cite{exp7} and $^{64}$Ni+$^{208}$Pb \cite{Ni2}. The isospin equilibration process in the MNT reactions can result in the production of very neutron-rich projectile-like fragments. A strong absorption of neutrons by the projectile was observed by the experiment \cite{exp7}. Such as that after neutron evaporation, $^{67}$Ni with production cross section about $15$ $\mu$b has been detected in the $\gamma-\gamma$ coincidence analysis.

\begin{center}
\textbf{IV. CONCLUSIONS }
\end{center}
In summary, the isotope production cross sections of $^{58}$Ni+$^{208}$Pb at $E_{\textrm{lab}}=328.4$ MeV have been calculated by the ImQMD model. The results show that the ImQMD model can successfully describe the MNT reactions at near-barrier energies. The isospin equilibration process between the $^{124}$Xe and $^{208}$Pb is studied. The isospin asymmetry in the inner of two reaction partners is uniform during the evolution of the system. The neutrons are preferably driven to the low density area. On the surface of the nuclei, the isospin asymmetry is larger than the inner of the nuclei. The direction of neutron flow is from the target to projectile along the low-density path of the neck during the reaction. When the projectile and target get close to each other, the neutron transfer path is firstly opened. The neutrons transfer is very sensitive to the symmetry energy coefficient. The neutron drift coefficient in the case of $C_s=32$ MeV is larger than the case of $C_s=28$ MeV with the growing up of the neck. The proton drift coefficient is weak correlation to the symmetry potential intensity. The isospin equilibration process in the reaction of $^{58}$Ni with $^{208}$Pb at $E_{\textrm{lab}}=345$ MeV is also investigated by the ImQMD model. It shows that $N/Z$ values of the products near the projectile increase rapidly with increasing mass transfer. In consequence, the very neutron-rich projectile-like fragments may be produced in the case of several nucleons transfer. The complete isospin equilibration events happen in symmetric fission-type reactions.

\begin{center}
\textbf{ACKNOWLEDGEMENTS}
\end{center}
This work was supported by the National Natural Science Foundation of China under Grants No. 11805015, No. 11635003, No. 11025524, No. 11161130520, and No. 11605270; the National Basic Research Program of China under Grant No. 2010CB832903; the European Commission's 7th Framework Programme (Fp7-PEOPLE-2010-IRSES) under Grant Agreement Project No. 269131; the Project funded by China Postdoctoral Science Foundation (Grant No. 2016M600956 and No. 2018T110069); the Beijing Postdoctoral Research Foundation (2017-zz-076).

\end{document}